\input harvmac
%
%
\noblackbox

\newcount\figno
\figno=0
\def\fig#1#2#3{
\par\begingroup\parindent=0pt\leftskip=1cm\rightskip=1cm\parindent=0pt
\baselineskip=11pt
\global\advance\figno by 1
\midinsert
\epsfxsize=#3
\centerline{\epsfbox{#2}}
\vskip 12pt
{\bf Fig.\ \the\figno: } #1\par
\endinsert\endgroup\par
}
\def\figlabel#1{\xdef#1{\the\figno}}
\def\encadremath#1{\vbox{\hrule\hbox{\vrule\kern8pt\vbox{\kern8pt
\hbox{$\displaystyle #1$}\kern8pt}
\kern8pt\vrule}\hrule}}

\def\frac#1#2{{#1 \over #2}}

\def\p{\partial}
\def\semi{\subset\kern-1em\times\;}
\def\bar#1{\overline{#1}}
\def\sqr#1#2{{\vcenter{\vbox{\hrule height.#2pt 
\hbox{\vrule width.#2pt height#1pt \kern#1pt \vrule width.#2pt}
\hrule height.#2pt}}}}
\def\square{\mathchoice\sqr64\sqr64\sqr{2.1}3\sqr{1.5}3}

                   \def\CL{{\cal L}}
                   
\def\CO{{\cal O}}

\def\p{\partial}

\def\ad{\bar a}

\def\Ab{\bar{A}}

\def\at{\tilde{a}}       
%

\Title{\vbox{\baselineskip12pt
\hbox{hep-th/0203221}
\hbox{UCLA-02-TEP-3}
\vskip-.5in}}
{\vbox{\centerline{Photons and Gravitons as Goldstone Bosons,}
\smallskip
\centerline{   and the Cosmological Constant}}}
\centerline{Per Kraus and E. T. Tomboulis}
\bigskip\medskip
\centerline{\it Department of Physics and Astronomy,}
\centerline{\it  UCLA, Los Angeles, CA 90095-1547,} 
\centerline{\tt pkraus, tombouli@physics.ucla.edu}
\medskip
\baselineskip18pt
\medskip\bigskip\medskip\bigskip\medskip
\baselineskip16pt
\noindent
We reexamine a scenario in which photons and gravitons arise as 
Goldstone bosons associated with the spontaneous breaking of 
Lorentz invariance.  We study the emergence of  Lorentz
invariant low energy physics in an effective field theory framework, 
with non-Lorentz invariant effects arising from radiative
corrections and higher order interactions.  Spontaneous breaking of the Lorentz
group also leads to additional exotic but weakly coupled Goldstone
bosons, whose dispersion relations we compute.  
 The usual cosmological constant problem is 
absent in this context: 
being a Goldstone boson, the graviton can never develop a 
potential, and the existence of a flat spacetime solution to the field
equations is guaranteed.       

\Date{March, 2002}
\lref\BjorkenVG{
J.~D.~Bjorken,
Annals Phys.\  {\bf 24}, 174 (1963).
}

\lref\BjorkenPE{
J.~Bjorken,
arXiv:hep-th/0111196.
}

\lref\ColladayFQ{
D.~Colladay and V.~A.~Kostelecky,
Phys.\ Rev.\ D {\bf 58}, 116002 (1998)
[arXiv:hep-ph/9809521].
}

\lref\BanksRH{
T.~Banks and A.~Zaks,
Nucl.\ Phys.\ B {\bf 184}, 303 (1981).
}

\lref\AtkatzRD{
D.~Atkatz,
Phys.\ Rev.\ D {\bf 17}, 1972 (1978).
}

\lref\Phillips{P. R. Phillips,
Phys. \ Rev. \ D {\bf 146}, 966, (1966).
}

\lref\OhanianQE{
H.~C.~Ohanian,
Phys.\ Rev.\  {\bf 184}, 1305 (1969).
}

\lref\Guralnik{
G. S. Guralnik,
Phys.\ Rev. D {\bf 136}, B1404 (1964).
}

\lref\WeinbergCP{
S.~Weinberg,
Rev.\ Mod.\ Phys.\  {\bf 61}, 1 (1989).
}

\lref\WeinbergKQ{
S.~Weinberg and E.~Witten,
Phys.\ Lett.\ B {\bf 96}, 59 (1980).
}

\lref\KachruHF{
S.~Kachru, M.~Schulz and E.~Silverstein,
Phys.\ Rev.\ D {\bf 62}, 045021 (2000)
[arXiv:hep-th/0001206].
}

\lref\ArkaniHamedEG{
N.~Arkani-Hamed, S.~Dimopoulos, N.~Kaloper and R.~Sundrum,
Phys.\ Lett.\ B {\bf 480}, 193 (2000)
[arXiv:hep-th/0001197].
}

\lref\CapperMV{
D.~M.~Capper and M.~J.~Duff,
Nucl.\ Phys.\ B {\bf 82}, 147 (1974).
}


\newsec{Introduction}

Massless particles can arise by a variety of mechanisms: gauge
symmetry, chiral symmetry, supersymmetry, and the spontaneous breaking
of global symmetry.   The masslessness of the photon and graviton is
typically associated with the existence of gauge symmetry;  here we 
explore an alternative option in which it is associated with the 
spontaneous breaking of Lorentz invariance, with the ``gauge'' fields
arising as Goldstone bosons.  This basic idea has a long history, dating
back to the 1964 work of Bjorken \BjorkenVG.  For related work see
 \refs{\Guralnik,  \Phillips,\OhanianQE,  \AtkatzRD}.  

At the level of low energy effective field theory, the usual argument
in favor of exact gauge invariance is based on Lorentz invariance.
Some form of gauge invariance is required in 
order to obtain an interacting, unitary, Lorentz invariant theory of  
massless particles with spin $1$ or $2$.   In a manifestly Lorentz invariant
formulation a violation of gauge invariance will typically imply a 
noncancellation of timelike and longitudinal modes, yielding a non-unitary
S-matrix in the physical sector.  In a noncovariant gauge fixed
formulation unitarity is manifest, but the action is required to
be formally gauge invariant in order to recover Lorentz invariance of the 
S-matrix.  

Especially in the case of gravity, there are various motivations for
exploring alternatives.  First, it is a basic fact that we inhabit
a universe that  is not Lorentz invariant at large scales due
to cosmological expansion.  Second, in the case of gravity gauge invariance
is insufficient to guarantee masslessness, since 
a potential $\Lambda^4 \sqrt{-g}$
is allowed.  Since a nonzero cosmological constant leads to a 
non-Lorentz invariant
vacuum, the original motivation for gauge invariance is not actually realized. 

As an alternative we turn to effective actions that yield
massless photons and gravitons as Goldstone bosons of spontaneously broken 
Lorentz invariance. 
Of course, the effective theory must be consistent with the observed accuracy
of Lorentz invariance at distances small compared to the curvature scale
of the universe.   This is achieved by effective theories consisting of 
three sorts of terms: gauge invariant kinetic terms, non-gauge invariant
potential terms, and small corrections to these.  
The simplest example is
${\cal L} = (F_{\mu\nu})^2 - V(A^\mu A_\mu)$.
As we will discuss, it is
not too hard to generate such effective actions from some more 
conventional underlying 
dynamics.  All terms in the action 
are taken to be Lorentz invariant; however the
 potential is assumed to give rise to a constant non-Lorentz
invariant vacuum expectation value. 
  The broken Lorentz generators imply
the existence of massless Goldstone bosons. 
In the absence of the
potential, the vacuum expectation value would have no physical effect,
being pure gauge; in particular exact Lorentz invariance would be maintained.
With the potential included gauge invariance is broken and with it exact
Lorentz invariance.
However, by definition the Goldstone bosons do
not appear in the potential; the expansion of the potential is in 
terms of massive fields.  Since Lorentz invariance is broken only by
the potential, the breaking will be suppressed by the inverse mass
of the heavy fields.  So at low energies we will have an approximately 
 Lorentz invariant
theory of Goldstone bosons.  The expansion of the potential to quadratic
order in fluctuations simply acts a gauge fixing term, so the low
energy theory is approximately that of photons or gravitons in a 
non-covariant gauge.  

Actually, there are additional interesting complications in the
quantum theory, where loops induce small non-Lorentz invariant kinetic
terms.  This leads to the appearance of additional non-Lorentz invariant
but ``weakly coupled'' Goldstone bosons whose effects we discuss. 
Directly related to this is that it is crucial to study whether the form
of our effective action is stable under
 radiative corrections, as these will generically 
induce all possible
operators consistent with the symmetries.  We study this question carefully,
and find that with reasonable starting assumptions the resulting low energy
physics appears approximately Lorentz invariant.  

Motivated by the cosmological constant problem, a multitude of authors have
experimented with modifying gravity in various ways (see \WeinbergCP\ for a 
review of some attempts).  From the point of 
view of low energy effective field theory the problem is that general
covariance allows one -- and only one -- potential term, 
$\Lambda^4 \sqrt{-g}$, and so unless this term vanishes there exists no 
solution to the equations of motion with constant fields.   But in a 
scenario in which the graviton is a Goldstone boson this problem does not
arise, since a Goldstone boson can never acquire a potential.  So even if some
scalar field undergoes a phase transition, contributing a term 
$V(\phi_0)\sqrt{-g}$ to the effective action, one knows that the vacuum
expectation value for massive fields can be shifted such that the Goldstone
boson gravitons remain massless.  Therefore, there will {\it always} exist an
exact 
vacuum solution with constant fields,  and on which  propagate
massless gravitons with approximately Lorentz invariant
physics.  This then guarantees the existence  of the sort of solution one
wants without fine tuning.   However, one should note that there may
also exist other solutions with space-time varying fields.   A complete
solution to the problem should address why a flat (or nearly flat) spacetime
solution is preferred; we discuss this in section 3.  

The first part of this paper is devoted to the detailed  study of the photon as
a Goldstone boson, but this is essentially a warmup for the more
interesting case of gravity, which provides our main motivation.
  There does not seem to be any obvious 
advantage in producing the photon as a Goldstone boson, and we have made
no attempt at a realistic model by including the other standard model fields.
We should also stress that we will work in an effective field theory 
framework in which the graviton is to be thought of as a composite of
more fundamental degrees of freedom.
 It may be worth mentioning why our scenario is not
in conflict with the  theorem of Weinberg and Witten \WeinbergKQ, 
which  rules out ``composite gravitons'' in a broad class of models.  
Specifically,
the theorem states that a Lorentz invariant theory with a Lorentz invariant
vacuum and a Lorentz covariant energy-momentum tensor cannot have a massless
spin two particle in its spectrum.  There is no conflict here since our vacuum
will not be Lorentz invariant.  

This paper is organized as follows.  In section 2 we study in detail the
example of the photon as a Goldstone boson. The photon example illustrates 
most of the important
general features and is much simpler computationally than the gravity
case. We also discuss
the relation to previous work on this subject.  In section 3 we
discuss the graviton and the cosmological constant problem in this context.
Section 4 has some final comments, and Appendix A contains technical
results for the gravity case.

\newsec{Photon as Goldstone Boson}

We begin by writing down a certain  effective action for a vector field $A_\mu$
coupled to matter. The motivation for this form will be discussed
subsequently.   The action is to be thought of as an effective action
defined at a UV cutoff scale $\Lambda$.  We thus consider the Lagrangian
\eqn\aa{ {\cal L} = N \left\{ -{1 \over 4}F^{\mu\nu}F_{\mu\nu} -V(A^\mu A_\mu)
 + {\rm higher~ derivatives}\right\} + {\cal L}_{\rm matter}(\phi,A_\mu)
+ { \cal O}(N^0).}
In the above, $N$ is some large number, and we have written out the 
leading $N$ terms in the action.  The action is Lorentz invariant --
all indices being raised and lowered with the Minkwoski metric -- but not
gauge invariant.  In particular, the potential $V(A^\mu A_\mu)$ is not gauge
invariant;  a crucial point is that this is the {\it only} non-gauge
invariant term at leading order in $N$. 
 In particular, the higher derivatives terms are terms
like $(F^{\mu\nu}F_{\mu\nu})^2$, $\partial^\alpha F^{\mu\nu}\partial_\alpha
F_{\mu\nu}$, etc.   Similarly, the generic matter fields
$\phi$  are gauge invariantly  coupled to $A_\mu$ in 
${\cal L}_{\rm matter}(\phi,A_\mu)$.  Apart from these stipulations the
action is generic in the sense that all dimensionless couplings (apart
from $N$) are of order unity;  that is, all dimensionful quantities are
of the order the cutoff $\Lambda$ to the appropriate power.  

Since in our effective action we have explicitly excluded various terms which
are consistent with the symmetries of the theory, two questions 
immediately come to the fore: how might an action of this form be 
generated from some underlying dynamics, and is the assumed structure
of the action stable under radiative corrections?   We now address
these two points in turn.

\subsec{Effective action from fermions}

To show why an action of the form \aa\ is fairly natural, we show how it
can be generated by integrating out some large number of fermion species.
This will be a generalization of the original mechanism proposed by Bjorken
\BjorkenVG, which considered four fermion interactions.   
So consider $N$ species of Dirac fermions $\psi_i$.  We imagine these
fermions being coupled to gauge fields which acquire masses
 at scale $\Lambda$.  Integrating out the massive gauge
bosons will yield an infinite set of fermion interactions, and we will
focus on the following subset:
\eqn\ab{ {\cal L}_\psi = \bar{\psi}_i (i\partial \!\!\!/ - m)\psi_i
+ N \sum_{n=1}^\infty \lambda_{2n} 
{\left(\bar{\psi}_i \gamma^\mu \psi_i\right)^{2n} \over N^{2n} }.}
Here, summation over flavor indices $i$ and spacetime indices $\mu$ is
implied.  We wrote the action to have a $U(N)$ flavor symmetry.  The couplings $\lambda_{2n}$ are
of order unity times the appropriate power of $\Lambda$,
\eqn\aba{\lambda_{2n} \sim \Lambda^{4-6n}.}  
Factors of 
$N$ have been inserted in order to give a well defined large $N$ limit.
In particular, the normalized bilinears 
${\cal O^\mu }  = {1 \over N} \bar{\psi}_i\gamma^\mu \psi_i$ 
then have correlators
scaling as $N^0$, and the action written in terms of  ${\cal O}$ has
an overall factor of $N$.  

We will employ the standard trick of rendering the action quadratic in fermions
by introducing an auxilliary field $A^\mu$.   We therefore consider
\eqn\ac{{ \cal L}_{\psi,A} = 
\bar{\psi}_i (i\partial \!\!\!/ -A \!\!\!/ - m)\psi_i - N V(A^\mu A_\mu).}
The potential $V$ is a power series in $A^\mu A_\mu$ with coefficients 
chosen such that by solving the algebraic equations of motion for $A^\mu$
and substituting back in we recover \ab.   The most familiar case
corresponds to a pure four fermion interaction, with only $\lambda_{2}$
nonvanishing, in which case $V(A^\mu A_\mu) = A^\mu A_\mu/4 \lambda_{2}$.
The quantum version of this theory is defined by a path integral (with a
cutoff) over the fields $\psi_i$ and $A^\mu$.  The idea is to imagine first
doing the path integral over $\psi_i$ to yield an effective action for
$A^\mu$.  Since $\psi_i$ is minimally coupled to $A^\mu$, provided we choose
a gauge invariant cutoff the terms in the effective action generated in this
way will be gauge invariant.  Furthermore, since there are $N$ species
of fermions the effective action will have an overall factor of $N$. 
Therefore, the form of the effective action is that of the first set of terms
in \aa.  

Now suppose we had included the other fermion terms in \ab\ that would 
certainly arise upon integrating out massive gauge fields.  This will
introduce other bilinears in the theory, {\it e.g.}, 
$\bar{\psi}_i \gamma^{\mu\nu} \psi_i$, 
$\bar{\psi}_i \gamma^{\mu}\partial^\nu  \psi_i$,  etc.  The above procedure
should then be generalized by introducing a new auxilliary field for each 
bilinear.  Integrating out the fermions then yields an effective action
for a set of interacting auxilliary fields.  The analysis rapidly
becomes complicated; however, one  expects on general grounds that
the auxilliary fields will acquire mass terms of order of the cutoff
and so can be neglected at lower energies.  By contrast, in our scenario 
certain components
of $A^\mu$ will remain massless since they will correspond to Goldstone bosons.

To reiterate somewhat, in our approach where we consider $A^\mu$ as the
only auxilliary field it is consistent to omit terms like 
$ (\bar{\psi}_i \gamma^{\mu} \psi_i)^2 
\bar{\psi}_j  \partial \!\!\!/ \psi_j$ which might seem to lead
to non-gauge invariant terms like 
$f(A^\mu A_\mu) F^{\alpha \beta}F_{\alpha \beta}$ in the
effective action for $A^\mu$.  If such fermionic terms are to be included
one should introduce a new scalar auxilliary field for the bilinear
$\bar{\psi}_i\partial \!\!\!/ \psi_j$, 
and we are not doing this for the reasons
stated above. 

We have thus demonstrated one possible way of generating an effective action 
with the
structure \aa, though there are presumably other ways as well.  For the 
most part we consider \aa\ in its own right, without reference to its 
origin. 

\subsec{Coupling to matter}

By coupling a matter field $\phi$ to the fermions via conserved currents,
$J^{\mu}(\phi)\bar{\psi}_i \gamma_{\mu} \psi_i$, we will generate the
matter couplings given in \aa.   An easy way to accomplish this is to
modify \ab\ by taking the mass for some of the fermions to be much less
than the cutoff $\Lambda$.  In this case we would  keep the light 
fermions in the low energy effective action rather than integrating 
them out.  From \ac\ we see that these fermions will be minimally coupled
to $A^\mu$.   Whether we use this or some other mechanism to generate the
matter couplings, it will be important that the matter action is gauge
invariant, at least up to the level of two derivative terms .

\subsec{Stability under radiative corrections}

Treating \aa\ as an effective field theory at scale $\Lambda$, to extract
out low energy physics we still need to integrate out the fluctuations
of $A^\mu$.   Since the potential term violates gauge invariance, once
we start computing loop diagrams all possible Lorentz invariant, but not
necessarily gauge invariant, terms will generically be generated.  Some
of these terms would lead, after spontaneous symmetry breaking,
 to large violations of Lorentz invariance at
low energies and so need to be suppressed.  The need to suppress such 
terms is our motivation for introducing the large number $N$.  Given the
form \aa, the loop expansion is an expansion in $1/N$, so the dangerous
terms will only arise at order $N^0$.   

So at order $N^0$ we need to consider all possible Lorentz invariant 
terms generated by computing loop diagrams for $A^\mu$.   At energies
low compared to the cutoff we can restrict attention to terms with 
at most two derivatives.  We further assume symmetry under charge 
conjugation $C$, acting
as sign reversal on $A^\mu$.  This forbids single derivative terms.
Terms with no derivatives just give a small correction to the potential in
\aa, which we are taking to be arbitrary,  so we need consider only
two derivative terms.  Up to integration by parts, there are seven 
independent terms: 

\eqn\ad{\eqalign{ 1) ~~&  f_1(A^2) \partial_\mu A_\nu \partial^\mu A^\nu \cr
2) ~~&  f_2(A^2) \partial_\mu A_\nu \partial^\nu A^\mu \cr
3) ~~&  f_3(A^2) A^\mu A^\alpha \partial_\mu A_\nu \partial_\alpha A^\nu \cr
4) ~~&  f_4(A^2) A^\nu A^\alpha \partial_\mu A_\nu \partial_\alpha A^\mu \cr
5) ~~&  f_5(A^2)A^\nu A^\alpha \partial_\mu A_\nu \partial^\mu A_\alpha \cr
6) ~~&  f_6(A^2) A^\mu A^\nu A^\alpha\partial_\mu \partial_\nu A_\alpha \cr
7) ~~&  f_7(A^2) A^\mu A^\nu A^\alpha A^\beta\partial_\mu  A_\nu 
 \partial_\alpha A_\beta \cr 
}}
Here $A^2 \equiv A^\mu A_\mu$.  As always, we assume that all dimensionful
couplings in $f_i$ are of order unity times the appropriate power of 
$\Lambda$.  As we will see, after spontaneous symmetry breaking some
of these terms will lead to low energy violations of Lorentz invariance
at order $1/N$.  

\subsec{Spontaneous symmetry breaking}

Our potential will generically have the form 
\eqn\ae{V(A^2) = \Lambda^4 \sum_{n=1}^\infty V_n \left( {A^2 \over \Lambda^2}
\right)^n,}
with the coefficients $V_n$ of order unity.   $V_n$ can be determined in
terms of the $\lambda_{2n}$ appearing in \ab.   We will assume that
the potential leads to spontaneous symmetry breaking,
\eqn\af{ \langle A_\mu \rangle = c \Lambda n_\mu,}
where for definiteness 
$n_\mu$ is a spacelike unit vector and $c$ is of order unity. 
This expectation value spontaneously breaks Lorentz invariance at the
cutoff scale $\Lambda$.  Nevertheless, we will see that low energy physics
is approximately Lorentz invariant.

We now expand the action around the vacuum \af\ by writing
\eqn\ag{A_\mu = c \Lambda n_\mu + a_\mu.}
To quadratic order in $a_\mu$  the potential becomes
\eqn\ah{ V = V(-c^2 \Lambda^2) + {1 \over 2 \alpha} (n \cdot a)^2 + \cdots,}
where 
\eqn\ai{ \alpha = {1 \over 4 V''(-c^2 \Lambda^2) } \sim {1 \over \Lambda^2}.}

Shifting the vacuum has a trivial effect on the $A^\mu$ kinetic terms
since, being gauge invariant, these depend only on derivatives of $A^\mu$.
Hence in the kinetic terms we can just replace 
$A^\mu \rightarrow a^\mu$.   We can similarly
make this replacement in the matter Lagrangian after performing a 
compensating gauge rotation of the matter fields:
\eqn\aj{{\cal L}_{\rm matter}(\phi,\langle A_\mu\rangle +a_\mu)=
{\cal L}_{\rm matter}(\phi',a_\mu),}
where $\phi'$ is a gauge transformation of $\phi$.  We will henceforth
drop the prime on $\phi$.  Therefore, the action takes the form
\eqn\ak{ {\cal L} = N \left\{ -{1 \over 4}F^{\mu\nu}F_{\mu\nu} 
-{1 \over 2 \alpha} (n \cdot a)^2
 + {\rm higher~ derivatives} + {\cal O}(a^3)\right\} 
+ {\cal L}_{\rm matter}(\phi,a_\mu)
+ { \cal O}(N^0).}

The $(n \cdot a)^2$ term plays the role of an axial gauge fixing term.
It gives a mass of order $\Lambda$ to a spacelike component of $a_\mu$.  
So to the above order our action takes the form of an axial gauge fixed
photon coupled gauge  invariantly to matter.  Neglecting the higher
derivative terms, the photon propagator is given by
\eqn\al{ -{i \over N p^2}\left(\eta_{\mu\nu} - {1 \over n\cdot p}
(n_\mu p_\nu + n_\nu p_\mu) - {p_\mu p_\nu \over (n\cdot p)^2}(\alpha p^2
- n^2) \right).}
As usual, only the first term contributes when the propagator is 
sandwiched between conserved current, and the result is Lorentz invariant. 
Corrections to this Lorentz invariant result are suppressed by 
$p/\Lambda$ and/or $1/N$, as will be discussed below.

\subsec{Goldstone bosons}

After spontaneous symmetry breaking we have ``massless particles''\foot{We 
employ quotation marks since we can no longer use the
standard definition of particles as being irreducible representations of
the (now spontaneously broken) Poincare group.  But it should be
clear what we mean when we use the particle terminology.} by virtue
of Goldstone's theorem.  In particular, a spacelike vector breaks 
the Lorentz group according to 
\eqn\am{ SO(3,1) \longrightarrow  SO(2,1).}
There are thus three broken generators corresponding to two rotations
and a boost.  The corresponding three Goldstone bosons are the three
components of $a^\mu$ orthogonal to $n^\mu$.  We choose a basis of
two transverse components and a timelike component:
\eqn\an{\eqalign{ &{\rm transverse}: ~~ \epsilon_\mu^{(1,2)}, \quad\quad
k\cdot \epsilon^{(1,2)} = n\cdot \epsilon^{(1,2)} = 0 \cr
&{\rm timelike}: ~~ \epsilon_\mu^{(0)} = k_\mu + (n\cdot k) n_\mu, \quad\quad
{\rm obeys:}~~ n\cdot \epsilon^{(0)} = 0.
}}
At low energies only the Goldstone bosons are relevant.  At leading order
in $N$ the gauge invariant form of the kinetic terms implies that 
only the transverse Goldstone bosons propagate, giving us the conventional
Lorentz invariant electrodynamics.  However, at order $N^0$ the timelike
component will also propagate, and this will lead to interesting effects. 

We now concentrate on the low energy physics of the Goldstone bosons coupled
to matter.   The Goldstone bosons can be thought of as coordinates on the
coset $SO(3,1)/SO(2,1)$.  Since the Goldstone bosons label flat directions
of the potential, the cubic and higher order terms from the expansion of
the potential all involve the massive component $n \cdot a$, and so are
irrelevant at low energies.  Now consider the order $N^0$ terms \ad.  
Note that after spontaneous symmetry breaking the terms (4) - (7) expanded
to quadratic order in fluctuations will all involve at least one factor
of $n \cdot a$.  Therefore, only terms (1) - (3) are relevant for
low energy physics.  Furthermore, one linear combination of (1) and 
(2) is proportional to  $(F_{\mu\nu})^2$ and so just provides a small 
correction to the order $N$ value of this term.  Hence we can omit one
linear combination, say (2), and focus only on terms (1) and (3).  
It is also convenient at this point to rescale $a^\mu \rightarrow 
a^\mu / \sqrt{N}$ to put the gauge kinetic term in standard form. 
Therefore  to order $N^0$, and discarding terms suppressed at low
energies by $p/\Lambda$, the effective action is
\eqn\ao{ {\cal L} =  -{1 \over 4}F^{\mu\nu}F_{\mu\nu} 
-{1 \over 2 \alpha} (n \cdot a)^2 + {1 \over 2}{c_1 \over N} \partial^\mu a^\nu
\partial_\mu a_\nu + {1 \over 2} {c_2 \over  N} n^\alpha n^\beta
\partial_\alpha a_\mu \partial_\beta a^\mu + 
{\cal L}_{\rm matter}(\phi,a_\mu /\sqrt{N}).}
Here we defined the order unity numerical coefficients $c_{1,2}$ as
\eqn\ap{f_1(\langle A \rangle ^2) = {1 \over 2} c_1, \quad
c^2 \Lambda^2 f_3(\langle A \rangle ^2) = {1 \over 2} c_2.}
We also kept the term $(n\cdot a)^2$ as a convenient way of implementing the
axial gauge condition.

\ao\ clearly shows the need for a $1/N$ suppression of the third and 
fourth terms in order to have approximate Lorentz invariance at low 
energies.  It is easiest to compute the propagator from the first
two terms and to think of the third and fourth terms as interactions. 
Then it is clear that when we compute interaction between conserved
matter currents we will get the standard QED results  at leading
order, with non-Lorentz invariant corrections occurring at order $1/N$. 

\subsec{Low energy spectrum}

We now determine the dispersion relations for the three Goldstone
bosons by solving the linearized equations of motion.  The latter are
\eqn\ap{ \left(1 + {c_1 \over N}\right) \partial^\mu \partial_\mu a^\nu
- \partial^\nu \partial_\mu a^\mu + { c_2 \over N }n^\alpha n^\beta
\partial_\alpha \partial_\beta a^\nu 
- {1 \over \alpha} n^\mu a_\mu n^\nu =0 .}

First consider the transverse modes.  Plugging in the ansatz (see \an)
\eqn\aq{a_\nu = \epsilon^{(1,2)}_\nu e^{-ik\cdot x}}
we find the dispersion relation
\eqn\ar{{\rm transverse:}~~ \left(1 + {c_1 \over N}\right)k^2 + {c_2 \over N}(n\cdot k)^2 =0.}
This corresponds to an anisotropic speed of light.  The speed of light 
parallel to $n^\mu$ differs from that orthogonal to $n^\mu$ by an
amount of order $1/\sqrt{N}$.  

Now consider the timelike mode.  We plug in the ansatz 
\eqn\as{a_\nu = (k_\nu + \gamma n_\nu)e^{-ik\cdot x}.}
We find
\eqn\at{\gamma = n\cdot k + {\cal O}(\alpha k^2),}
and the dispersion relation
\eqn\au{{\rm timelike:}~~k^2 - {N \over c_1} (n \cdot k)^2 = 0.}
In the dispersion relation we have dropped terms down by $1/N$ or $\alpha k^2$.
The dispersion relation \au\ is non-Lorentz invariant at leading order.
The timelike modes propagate at the ordinary speed of light in directions
orthogonal to $n^\mu$, but at a speed of order $\sqrt{N}$ in the direction
parallel to $n^\mu$.  We are assuming that $c_1 >0$.  

The physics of the transverse modes is thus standard up to small corrections,
while that of the timelike mode is quite exotic.  The reason for this
is that the timelike mode does not propagate with respect to the leading
$N$ gauge invariant kinetic terms of \aa.  It only acquires a kinetic
term at order $N^0$, and these terms are non-Lorentz invariant after
spontaneous symmetry breaking.  

While exotic, the timelike mode leads to acceptably 
small effects for sufficiently
large $N$.  Its contribution to the interaction between conserved
currents is suppressed by $1/N$, since as we have discussed we can 
use the standard axial gauge propagator at leading order and regard
corrections as coming from interaction vertices with coefficients 
of order $1/N$.  

The timelike mode is also suppressed by phase space considerations.
For fixed available energy $k^0$ the dispersion relation forces
\eqn\av{ |k \cdot n|  < \sqrt{{c_1 \over N}} k^0.}
Consider putting the system in a box of size $L$.   Then for 
$N > (k^0 L)^2$ only the zero momentum mode parallel to $n^\mu$ survives,
yielding a phase space suppression proportional to $1/L$.  

While the timelike mode gives small corrections to the interaction between
conserved currents it could have more dramatic consequences given
its unusual dispersion relation.  We should emphasize that the result \au\
does not necessarily imply faster than light\foot{Here we refer
to the speed of light as the speed of the transverse modes after
spontaneous symmetry breaking.} signal propagation, since \au\
is only valid for long wavelengths.  Also, even if \au\ could be 
extrapolated to short wavelengths so that signals could propagate
faster than light, there would be no conflict with causality since 
Lorentz invariance has been spontaneously broken by a preferred frame.
It would be interesting to study the physics of the timelike mode in
more detail.

\subsec{Summary and relation to previous work}

Let us summarize what has been accomplished.  We have shown that 
spontaneous breaking of Lorentz symmetry can lead to an approximately
Lorentz invariant low energy theory of massless photons coupled to matter. 
This is possible in the context of a theory in which gauge 
invariance is violated at leading order only by a potential term.  
Since the Goldstone boson photons do not appear in the potential, the 
Lorentz violating condensate leads to only small corrections to the
low energy physics of the photons.  On the other hand, the existence
of three broken Lorentz generators implies the existence of a third
Goldstone boson whose physics is not even approximately Lorentz invariant.
However, its effects are suppressed by $1/N$ and phase space considerations.
Altogether, Lorentz invariance appears as an approximate symmetry of the
low energy world.

This is a good place to compare and contrast with previous work on this
subject, in particular the original work of Bjorken \BjorkenVG.  
The main difference
is that we have taken a modern effective field theory point of view, 
emphasized that the violation of Lorentz invariance is real, and pointed
out the existence of an extra Goldstone boson.  Earlier work started from 
a four-fermi interaction, {\it i.e.} just keeping the term $\lambda_2$ is
\ab.  The trouble with this is that it is incompatible with spontaneous
symmetry breaking, since it corresponds to a potential $V \sim A^\mu A_\mu$
with no higher order terms.  A reflection of this is that the condensate was
never actually computed in earlier work, but was either {\it assumed} to arise
somehow, or emerged after formal manipulations with divergent integrals.  
The claim was then made that the physics after spontaneous symmetry breaking
was the usual {\it exactly} Lorentz invariant quantum electrodynamics. 
This conclusion was again dependent on manipulating divergent quantities.
As far as we can tell, the origin of this claim is that if one takes the
pure four fermi interaction and assumes (wrongly) that this leads to 
spontaneous symmetry breaking, then expanding $V \sim A^\mu A_\mu$ 
around the new vacuum would yield an axial gauge fixing term and nothing more.
This of course gives   usual QED in axial gauge. But from our point of view it 
is clear what would actually happen in this theory.  The four fermi
theory either leads to an instability or to a stable vacuum with a massive
vector field $A_\mu$.  In neither case does one find QED.   On the other
hand, the problem disappears once one includes the higher order fermion
terms as we have done; this is also the natural starting point from the
view of effective field theory. 

Since earlier work took the point of view that the spontaneous breaking
of Lorentz invariance was somehow fictitious, the existence of the 
extra Goldstone boson was not noted.  

Before turning to gravity we should also note that some of the above
criticisms were commented on recently by Bjorken \BjorkenPE.  
In particular it was noted
that the four fermi theory by itself is inadequate, and that some real
violations of Lorentz invariance should be expected once quantum effects
are taken into account.  These points were also examined by Banks and Zaks
in  \BanksRH\
in the context of non-abelian gauge fields; they also concluded that there
is a real violation of Lorentz invariance.  
We hope to have resolved these issues here.

\newsec{Graviton as Goldstone boson}

We now turn to our main interest: producing a graviton as a Goldstone
boson.  Fortunately, the analysis closely parallels the photon case,
and so we can draw on our experience from that example to navigate in the
more complicated gravitational setting.  The main difference is in the
different pattern of spontaneous Lorentz breaking as well as in the
connection to the cosmological constant problem.  The cosmological 
constant provides a motivation for modifying the low energy effective
theory of general relativity, and indeed we will see that the problem
is avoided in the sense that the Goldstone boson graviton remains massless
even in the presence of vacuum energy.

To adapt the previous approach to the case of gravity we consider
an effective action in direct 
analogy with \aa,
\eqn\ba{{\cal L} = N \left\{ \Lambda^2 \sqrt{-g}R(g)  -\Lambda^4 V(h)
 + {\rm higher~ derivatives}\right\} + {\cal L}_{\rm matter}(\phi,g)
+ { \cal O}(N^0).}
Here $h$ is defined via expansion of the metric around flat spacetime
\eqn\bb{g_{\mu\nu} =\eta_{\mu\nu} + h_{\mu\nu}.}
Flat spacetime  thus plays a preferred role in this context; indeed we
think of the underlying dynamics as that of a nongravitational theory
in Minkowski spacetime.  Appendix A discusses one possibilty for such
an underlying theory.  
The Einstein-Hilbert term in \ba\ is standard while the potential is some
generic Lorentz invariant function of $h_{\mu\nu}$ with indices contracted
with $\eta_{\mu\nu}$.  Odd powers of $h$ are allowed, for instance
the term $h^\mu_\mu$ can appear in the expansion of $V(h)$.  As in the
photon case the higher derivative terms are generally covariant, as is
the matter action.  General covariance is violated only by the potential. 

Note that the observed Newton's constant will be
\eqn\bba{ G_N \sim {1 \over N \Lambda^2}.}
Therefore, the cutoff $\Lambda$ is smaller by a factor of $1/\sqrt{N}$
compared to the usual Planck scale.  Indeed, this lowered value of the
cutoff is responsible for suppressing loop corrections.  $N$ is a large
number but presumably need not be more than $10^4$ or so.   Thus the
cutoff can still be well above observeable energy scales.   

We consider potentials leading to a vacuum expectation value for $h_{\mu\nu}$.
By performing a Lorentz transformation we can bring the expectation value
to the form
\eqn\bc{ \langle h_{\mu\nu} \rangle = \pmatrix{
\bar{h}_{00} &  &  & \cr
& \bar{h}_{11} && \cr
&& \bar{h}_{22} & \cr
&&&\bar{h}_{33} }.}
For $\bar{h}_{\mu\mu}$ all nonvanishing and distinct, the Lorentz group will
be completely broken:
\eqn\bd{SO(3,1) \longrightarrow {\rm nothing}.}  
Being dimensionless, we expect $\bar{h}_{\mu\mu}$ of order unity. 
Therefore there will be $6$ Goldstone bosons corresponding to the six
broken Lorentz generators.  The Lorentz generator $J^{\mu\nu}$ acts on
$\langle h_{\mu\nu} \rangle$ by exciting the $\mu\neq \nu$  
components.  So the Goldstone bosons are the six off-diagonal components
of the symmetric matrix $h_{\mu\nu}$.  Fluctuations of the diagonal 
components will generically correspond to massive fields.   We will
ultimately associate two of the Goldstone bosons with the two physical
polarizations of the graviton, while the remaining four will appear
in analogy with the timelike Goldstone boson in the the photon case.  

We now consider fluctuations around the vacuum by writing
\eqn\be{ h_{\mu\nu} = \langle h_{\mu\nu}\rangle + \tilde{h}_{\mu\nu}.}
The expansion of the potential will correspond to mass terms for
the diagonal components of $\tilde{h}_{\mu\nu}$.   The precise form
of this mass matrix is not important, so we will
write
\eqn\bfa{ V(h) = {\rm constant} + \Lambda^4 \sum_{\alpha=0}^{3}
( f_\alpha n_{(\alpha)}^\mu n_{(\alpha)}^\nu \tilde{h}_{\mu\nu})^2 + {\cal O}
(\tilde{h}^3).}
Here 
\eqn\bg{n_{(\alpha)}^\mu  = \delta^\mu_\alpha,}
and $f_\alpha$ are numbers of order unity.  

We can simplify the action by performing a general coordinate transformation
to put the background metric back in standard form.  In particular,
introduce new coordinates $x'^\mu$ such that
\eqn\bh{ {\partial x'^\mu \over \partial x^\alpha}
{\partial x'^\nu \over \partial x^\beta }\eta_{\mu\nu} = \eta_{\alpha \beta} + 
\langle h_{\alpha \beta} \rangle.}  
The metric appearing in the action will then be 
$\eta_{\mu\nu} + \tilde{h}'_{\mu\nu}$ where
\eqn\bi{{\partial x'^\mu \over \partial x^\alpha}
{\partial x'^\nu \over \partial x^\beta }\tilde{h}'_{\mu\nu}(x')=
\tilde{h} _{\alpha\beta}(x).}
We similarly act with a coordinate transformation on the matter fields;
{\it e.g.} for a scalar
\eqn\bj{ \phi'(x') = \phi(x).}
After changing the integration variable to $x'$ the potential term 
is modified while the general covariant terms are of course invariant.
Given the form \bi, the modification of the potential can be absorbed
in a redefinition of the constants  $f_\alpha$.  So after the 
coordinate transformation our action takes the form 
\eqn\bk{{\cal L} = N \left\{ \Lambda^2 \sqrt{-g}R(g)  
-\Lambda^4 V(\tilde{h}')
 + {\rm higher~ derivatives}\right\} + {\cal L}_{\rm matter}(\phi',g)
+ { \cal O}(N^0),}
with 
\eqn\bl{ g_{\mu\nu} = \eta_{\mu\nu} + \tilde{h}'_{\mu\nu}}
and 
\eqn\bm{ V(\tilde{h}') = {\rm constant} + \Lambda^4 \sum_{\alpha=0}^{3}
( f_\alpha n_{(\alpha)}^\mu n_{(\alpha)}^\nu \tilde{h}'_{\mu\nu})^2 + {\cal O}
(\tilde{h}'^3).}
We henceforth relabel fields: $\tilde{h}' \rightarrow h$, $\phi'
\rightarrow \phi$.  

We think of the quadratic terms in the potential as gauge fixing terms,
corresponding to the gauge
\eqn\bn{ g_{\mu\mu} = \eta_{\mu\mu}, \quad {\rm no ~ sum}.}
This defines an acceptable noncovariant gauge.
The graviton propagator in this gauge is extremely complicated and
unwieldy, and so we will not display it here.  Fortunately, for leading
order calculations we only need to know that it has the structure of
the standard covariant propagator
\eqn\bo{ {-i \over p^2} \left( \eta_{\mu\alpha} \eta_{\nu\beta}
+\eta_{\mu\beta} \eta_{\nu\alpha} - \eta_{\mu\nu} \eta_{\alpha \beta}
\right)}
plus terms with at least one factor of $p$ (vector) in the numerator.  
Sandwiched between conserved energy momentum tensors the latter $p$
terms vanish, and so we recover 
the standard Lorentz invariant result.  Of course this is no
suprise, since \bn\ represents a valid gauge choice.  

The low energy physics is therefore quite similar to what we found in
the photon example.  We will have two graviton states which propagate
at the speed of light, up to a small anisotropic correction.  Further
there are four additional Goldstone bosons that acquire kinetic terms
at order $1/N$.  These will have highly non-Lorentz invariant 
dispersion relations, but their couplings to conserved currents are
suppressed by $1/N$.   Working out these dispersion relations explicitly
would be quite involved given the large number of terms in the action
at order $1/N$ and the proliferating indices.  However from our discussion
of the photon example it should be clear that the essential physics is
independent of these details.

\subsec{The cosmological constant}

Notice that we have obtained an approximately Lorentz invariant theory
of gravity without making any specific assumptions about the form
of the potential $V(h)$.  Therefore, we see that if the potential is
suddenly modified, say by a matter phase transition, then the vacuum
expectation value of $h$ can simply shift to the new minimum, leaving us
again with an approximately Lorentz invariant theory.  In particular,
the term $\sqrt{-g} V_{\rm matter}(\phi_0)$ can  be added to our previous 
potential and the analysis proceeds as before.  We have therefore 
evaded the usual cosmological constant problem.   The usual problem 
arises because of general covariance: only a single potential term
is allowed, $\Lambda^4 \sqrt{-g}$, and a nonzero value of this term
is incompatible with a Lorentz invariant solution.  If one is willing 
to violate general covariance by writing a more general  potential
then this conclusion need not follow.  Indeed, if the graviton is 
a Goldstone boson one is guaranteed to find a solution with constant 
fields and a massless graviton.  What is perhaps surprising is that the 
physics around such a non-Lorentz invariant solution is
approximately Lorentz invariant, as we have seen. 

On the other hand, the above discussion does not immediately imply
that the approximately Lorentz invariant solutions are the {\it only} 
solutions.  Indeed, at least in the weak field approximation we 
will find additional approximately de Sitter and anti-de Sitter
solutions.  This follows from the fact that at low energies and for
weak fields our theory is that of standard gravity in a noncovariant
gauge, plus additional weakly coupled Goldstone bosons.    If one now
takes a standard solution with a given energy momentum tensor and expresses
it in our gauge, this must continue to be an approximate solution of our 
theory.  The existence of multiple solutions is hardly surprising 
given that we are solving (at leading order) second order differential
equations\foot{It is also reminiscent of  brane world scenarios
for addressing the cosmological constant problem 
\refs{\KachruHF, \ArkaniHamedEG}. But there the 
Lorentz invariant solution on the brane is tied up with
a naked singularity away from the brane, and so need not exist.}.
Note that the approximately Lorentz invariant solution with
constant fields is guaranteed to be an exact solution as it just
corresponds to extremizing the potential, while the other solutions with
spacetime varying fields need
not be exact.  In a more conventional field theory context one would
expect a solution with time dependent fields to eventually settle down
to a static solution by radiating energy.  One might expect the same here,
with the time dependent fields of the de Sitter type solutions 
radiating away leading to the solution with constant fields.  
A realistic proposal in this framework must involve showing
how to make the transition from an expanding radiation or 
matter dominated universe to the approximately Lorentz invariant solution
discussed above.
We hope to return to this in future work.

\newsec{Concluding Remarks}

Building on the work of Bjorken,
we have shown that massless photons and gravitons can be produced as 
Goldstone bosons associated with the spontaneous breaking of Lorentz
invariance, and with low energy physics appearing Lorentz invariant to 
high accuracy.   The most dramatic effect of the Lorentz breaking is the
existence of additional weakly coupled Goldstone bosons obeying highly 
non-Lorentz invariant dispersion relations.  These fields would be 
difficult to detect as they couply weakly to conserved currents.  
A rather general framework for studying Lorentz violating extensions
of the Standard Model has been developed (see, {\it e.g.}, \ColladayFQ),
and it might be useful to study some of our results in that language.  

We find it interesting that the observed low energy physics of gravity
can be produced in the context of an effective field theory that differs
markedly from general relativity, and which does not suffer from the
usual cosmological constant problem.  While it remains to be seen whether
a theory of this type could be incorporated into a truly fundamental 
framework or be developed into a realistic cosmology, it seems to be an
idea worth pursuing.  

\appendix{A}{}
In this appendix we indicate one way the gravitational 
effective action \ba\ may emerge from some underlying conventional 
dynamics. There may be others. 
Here we employ the same mechanism as in the photon case, \aa.    
Thus we consider $N$ fermions coupled to gauge fields that acquire masses. 
We then imagine integrating out the fields from some initial 
$\Lambda_0$ down to a scale 
$\Lambda$ obtaining the effective action 
\eqn\Aa{
\bar{\psi}_i\,\left(\,i\not\! \partial - M\,\right)\,\psi_i 
+ 4\pi^2 N \sum_k\;C_k(\Lambda_0, 
\Lambda)\,\CO_k(\Lambda)} 
involving an infinite set of fermion 
interactions. Restricting first to the subset consisting only of powers of 
\eqn\Ab{
\CO_{\mu\nu} = {1\over N}\, \bar{\psi}_i \,{i\over 2}\left(\,\gamma_{\mu} 
\buildrel\rightarrow\over\partial_\nu  - 
\gamma_{\mu}\buildrel\leftarrow\over\partial_\nu
\,\right)\psi_i\;,} 
we introduce the symmetric auxiliary field $h^{\mu\nu}$ to render the 
effective action quadratic in the fermions, so that we may write it  
in the form (cp. \ac): 
\eqn\Ac{
\CL_{\psi,h} =  (\eta^{\mu\nu} + h^{\mu\nu})\,\bar{\psi}_i
\,{i\over 2}\left(\,\gamma_{\mu} 
\buildrel\rightarrow\over\partial_\nu  - 
\gamma_{\mu}\buildrel\leftarrow\over\partial_\nu
\,\right)\psi_i 
-M\bar{\psi}_i\psi_i -N{\Lambda^4\over 4\pi^2}\,V(h) \;.
}
All indices are raised and lowered by the flat metric $\eta_{\mu\nu}$.

Integrating out the fermions, the effective action from the 
resulting determinant can be expressed, as usual, as the 
sum over all fermion 1-loop diagrams with  external 
$h$ legs. Note, in particular, that the diagram with one 
external $h$-leg is in general nonvanishing. This 
reflects the fact that  
\Ab\ has a nonvanishing expectation (proportional to 
$\eta_{\mu\nu}$) even in ordinary perturbation theory on a  
Lorentz invariant vacuum, i.e. interactions built from \Ab\ 
shift the classical background. Correspondingly,    
all terms, including a linear term, are included in the general 
potential $V(h)$ in \ba\ (cp. discussion in the text). 

Explicit evaluation of the fermion-loop graphs with one and two 
external $h$-legs gives, after a lengthy computation, the 
contribution to the effective action to $\CO (h^2)$: 
\eqn\Ad{\eqalign{
\CL^{(2)} & =   N I_4\,\Big[\,-h^\mu_\mu + {1\over 2}h^{\mu\nu}h_{\mu\nu} 
+ {1\over 2} (h^\mu_\mu)^2\,\Big] \cr
  & +     {N \over 6}\,I_2\,\Big[\, (\p_\lambda h^{\mu\nu})^2 
- (\p_\nu h^\mu_\mu)^2 
+ 2\,\p_\mu h^{\mu\nu}\p_\nu h^\lambda_\lambda - 2(\p_\mu h^{\mu\nu})^2 
\,\Big] \cr 
 & -   {N\over 20}\, I_0\,\Big[\, (\square \;h^{\mu\nu})^2 - \frac{1}{3}  
(\square \; h^\mu_\mu)^2 - 2 (\p_\mu \p_\lambda h^{\lambda\nu})^2 
+\frac{2}{3}\, \p_\mu \p_\nu h^{\mu\nu}\square\; h^\lambda_\lambda \cr 
&  \qquad \quad +   \frac{2}{3}\,(\p_\mu \p_\nu h^{\mu\nu})^2 \,\Big]  
+ \cdots \cr}
}
\Ad\ displays explicitly the leading terms, i.e. the local, cut-off 
dependent (`divergent') part of the result of the loop integrations.  
The ellipses denote the subleading pieces from the finite, 
non-local parts, which can be expanded in powers of $\square\,/M^2$, and   
contribute to higher derivative interactions relevant only for 
short distance behavior near the cutoff. Dimensional regularization, 
under the usual correspondence $\ln\,\Lambda \leftrightarrow 
(\,1/\epsilon + \hbox{const.})$, gives
\eqn\Ae{
I_n = {1\over (4\pi)^2}\,M^n\,\ln(\frac{\Lambda^2}{M^2})\;,\qquad 
n=0,2,4 \,.
}
In Pauli-Villars regularization, which appears more physical in the 
present context, one has  
\eqn\Af{ 
I_n = {1\over (4\pi)^2}\,\sum_{k=1}^{3} c_kM_k^n\,\ln(\frac{M_k^2}{M^2})\;, 
\qquad n=0,2,4 \,.
}
Three regulator masses $M_k$, of order of the cutoff $\Lambda$,  
are required here with coefficients $c_k$ 
satisfying $\sum_{k=1}^{3} c_kM_k^n + M^n=0$, for $n=0,2,4$. 

\Ad\ is now seen to be 
the flat space expansion 
to 2nd order of the gravitational action
\eqn\Ag{
\CL = \sqrt{-g}\,N\,\Big[\,I_4 + {1\over 6}I_2\,R - {1\over 20}I_0\,
(\,R_{\mu\nu}R^{\mu\nu} - \frac{1}{3}R^2)\,\Big] 
}  
with the metric expressed in terms of the vierbein\foot{Curved and 
flat indices are denoted by Greek and Latin letters, respectively.} 
\eqn\Ah{ 
g^{\mu\nu} = e^\mu_ae^{a\nu}\;, \qquad 
e^\mu_a = \delta^\mu_a + h^\mu_a \;, 
} 
and 
\eqn\Ai{
h^{\mu\nu} = \delta^\mu_a\,h^{a\nu}\,. 
}
Taking $h^{\mu\nu}$ to be symmetric, as done in the above calculation, 
amounts to the (standard) local Lorentz gauge fixing  
to a symmetric vierbein. 
(It is known that the antisymmetric part in fact decouples in 
\Ag.) Thus our effective gravitational action 
\ba\ is reproduced to this order. 

To see how this comes about, note that the result \Ad\ is 
precisely what one obtains after integrating out the fermion fields 
in the Lagrangian for $N$ fermions in curved space:
\eqn\Aii{
\CL= e^{(1-2w)}\Big[\,e^\mu_a\,\bar{\psi}_i{i\over 2}\left(\, \gamma^a  
\buildrel \rightarrow \over \nabla_\mu - 
\buildrel \leftarrow \over \nabla_\mu
\gamma^a\,\right)\psi_i - M\bar{\psi}_i\psi_i\,\Big] \,,
}
to 2nd order in the expansion about 
flat space \Ah. In \Aii,   
$e=\hbox{det}\,e_{a\mu}$, and 
\eqn\Aiii{
\buildrel{\buildrel \rightarrow \over\leftarrow}\over \nabla_\mu
\equiv\buildrel{\buildrel\rightarrow\over\leftarrow}\over \partial_\mu 
\mp \frac{i}{2}\omega_{\mu\,ab} S^{ab} 
}
with Lorentz generators $S^{ab}={i\over 4}[\gamma^a,\gamma^b]$. 
The spinor $\psi$ may, in general, be taken to transform under 
general coordinate transformations as a density of weight $w$. 
The result of the explicit computation is in fact found to be 
independent of $w$ (see, for
example, \CapperMV).   Indeed, note that, in terms of 
the functional integration, the 
factor $e^{(1-2w)}$ can be absorbed by the change of variables: 
$\psi \to e^{(1-2w)/2}\psi^{\;\prime}$, $\bar{\psi} \to e^{(1-2w)/2}
\bar{\psi}^\prime$.   
The spin connection $\omega_{\mu\,ab}$ is given in terms of the 
vierbein by 
\eqn\Aiv{\eqalign{ 
\omega_{\mu\,ab}&= \frac{1}{2}e^a_\mu\,(\,T_{amn} - T_{mna}- T_{nam}\,)
\;,\cr
T^a_{mn} & =  (\,e^\rho_m e^\sigma_n - e^\rho_n e^\sigma_m\,) 
\partial_\sigma e^a_\rho \,. \cr} 
}
\Aiv\ implies that the connection terms in \Aii, 
in the expansion \Ah,  
do not give a $h\bar{\psi}\psi$ vertex, but only 
`seagull' $h^2\bar{\psi}\psi$ and higher $h$ powers vertices.  
This is most easily seen by 
rewriting the connection terms in 
\Aii, after a little rearangemant, in the form 
$$e^\mu_m \bar{\psi}\,{1\over 4}\omega_{\mu\,ab}\,\epsilon^{mabc}\gamma_c
\gamma^5\psi\;.$$ 
Furthermore, it follows from this form that all fermion 1-loop diagrams with  
two external $h$-legs do not receive any contribution from 
the spin connection interaction. The result to $\CO(h^2)$ 
thus agrees with that 
obtained from \Ac.

The vertices from the spin connection terms will, however, 
contribute to the diagrams with three or more external legs, 
reproducing \Ag\ to all orders, as dictated by the 
general coordinate invariance of \Aii. It may therefore   
appear that, in addition to \Ab, one would need 
an infinite set of different operators\foot{Note that 
\Aiv\ contains both $e^{a\mu}$ and its inverse.} 
from \Aa, with precisely specified coefficients, 
to be included in \Ac\ in order to generate \Ag.  
This, however, is not the case. In the context of \Aa,  
the connection arises  
naturally when \Ac\ is extended to include 
the set of powers of the operator 
\eqn\Av{
\CO_\mu^{\kappa\lambda} = {i\over 4N}\,\bar{\psi}_i \,\{\gamma_\mu, 
\,[\gamma^\kappa,\gamma^\lambda]\}\psi_i \;, 
}
in addition to those of \Ab. Introducing the corresponding 
independent auxiliary field $\omega^\mu_{\kappa\lambda}$,  
\Ac\ is extended to    
\eqn\Avi{\eqalign{
\CL_{\psi,h}& =  (\eta^{\mu\nu} + h^{\mu\nu})\,\bar{\psi}_i
{i\over 2}\left(\,\gamma_\mu 
\buildrel\rightarrow \over\partial_\nu  - 
\gamma_\mu \buildrel \leftarrow \over\partial_\nu
 + \frac{1}{8}\,\omega_{\nu\,\kappa\lambda}\{\gamma_\mu, 
\,[\gamma^\kappa,\gamma^\lambda]\}
\,\right)\psi_i \cr
&  -M\bar{\psi}_i\psi_i -N{\Lambda^4\over 4\pi^2}\,V(h, \omega) 
\;.\cr }
}
The fermionic part of \Avi\ is equivalent to 
\Aii\ in the first order (Palatini) formulation. 
  
Now, in the first order formalism, the nonpropagating 
connection in \Aii\  
serves as a constraint field enforcing vanishing of 
torsion $e^\mu_m\CO^{m\,ab}$ generated by the fermions.

\Avi\ differs from \Aii\ by the presence of a potential in 
$\omega$. Hence, variation of the connection will not imply 
vanishing torsion, and $\omega$ will not be expressible 
entirely in terms of the vierbein as in 
\Aiv.\foot{This is analogous to considering 
\Aii\ in the first order formalism not just by itself, 
but with the addition of the Einstein-Hilbert  
action. The latter  
provides a potential for the connection $\omega$, thus leading to 
the usual result of torsion generated by fermions. 
This is of course the usual situation. In the above we were lead to 
consider \Aii\ by itself in the context of generating the 
Einstein-Hilbert action from \Aa.} 
For illustration purposes, we may adopt a model where the quadratic 
terms in the potential for $\omega$ are supressed. Then, upon 
integrating out the fermions, \Ag\ and the effective 
gravitational action \ba\ are reproduced to within 
small deviations. This is easily seen to be stable under 
radiative corrections from graviton loops. 

It is perhaps worth pointing out again that, as we saw, just the fermionic 
self-interactions of products of \Ab\ in flat space-time   
already suffice to fully reproduce to second order 
the Einstein-Hilbert (plus $R^2$ 
terms) parts in our 
effective gravitational action \ba, i.e 
reproduce the full content of linearized GR.

\bigskip\medskip\noindent
{\bf Acknowledgements:}
This work was supported  by NSF grants PHY-0099590 and PHY-9819686.

\listrefs

\end